\documentclass[sigconf]{acmart}

\AtBeginDocument{%
  \providecommand\BibTeX{{%
    \normalfont B\kern-0.5em{\scshape i\kern-0.25em b}\kern-0.8em\TeX}}}
    
\usepackage{xcolor}
\usepackage[normalem]{ulem}
\usepackage{listings}

\begin{document}

\title{Engineering for a Science-Centric Experimentation Platform}

\author{Nikos Diamantopoulos}
\email{ndiamantopoulos@netflix.com}
\affiliation{%
  \institution{Netflix, Inc.}
  \city{Los Gatos}
  \state{California}
  \country{USA}
}

\author{Jeffrey Wong}
\email{jeffreyw@netflix.com}
\affiliation{%
  \institution{Netflix, Inc.}
  \city{Los Gatos}
  \state{California}
  \country{USA}
}

\author{David Issa Mattos}
\email{davidis@chalmers.se}
\affiliation{%
  \institution{Chalmers University of Technology}
  \city{Gothenburg}
  \country{Sweden}
}

\author{Ilias Gerostathopoulos}
\email{gerostat@in.tum.de}
\affiliation{%
  \institution{Technical University of Munich}
  \city{Munich}
  \country{Germany}
}

\author{Matthew Wardrop}
\email{mawardrop@netflix.com}
\affiliation{%
  \institution{Netflix, Inc.}
  \city{Los Gatos}
  \state{California}
  \country{USA}
}

\author{Tobias Mao}
\email{tmao@netflix.com}
\affiliation{%
  \institution{Netflix, Inc.}
  \city{Los Gatos}
  \state{California}
  \country{USA}
}

\author{Colin McFarland}
\email{cmcfarland@netflix.com}
\affiliation{%
  \institution{Netflix, Inc.}
  \city{Los Gatos}
  \state{California}
  \country{USA}
}

\renewcommand{\shortauthors}{Diamantopoulos, et al.}

\begin{abstract}
Netflix is an internet entertainment service that routinely employs experimentation to guide strategy around product innovations. As Netflix grew, it had the opportunity to explore increasingly specialized improvements to its service, which generated demand for deeper analyses supported by richer metrics and powered by more diverse statistical methodologies. To facilitate this, and more fully harness the skill sets of both engineering and data science, Netflix engineers created a science-centric experimentation platform that leverages the expertise of data scientists from a wide range of backgrounds by allowing them to make direct code contributions in the languages used by scientists (Python and R). Moreover, the same code that runs in production is able to be run locally, making it straightforward to explore and graduate both metrics and causal inference methodologies directly into production services.

In this paper, we utilize a case-study research method to provide two main contributions. Firstly, we report on the architecture of this platform, with a special emphasis on its novel aspects: how it 
supports science-centric end-to-end workflows without compromising engineering requirements.
Secondly, we describe its approach to causal inference, which leverages the potential outcomes conceptual framework to provide a unified abstraction layer for arbitrary statistical models and methodologies.

\end{abstract}


\settopmatter{printacmref=false}  
\renewcommand\footnotetextcopyrightpermission[1]{} 
\pagestyle{plain} 

\maketitle

\section{Introduction}

Understanding the causal effects of product and business decisions via experimentation is a key enabler for innovation and improvement~\cite{Kohavi2009,Olsson2014,fabijan_evolution_2017}, and the gold-standard of experimentation is the randomized controlled trial design (also known as A/B testing) \cite{box_statistics_2005,xu_infrastructure_2015,xie_improving_2016}.

In this paper, we will be presenting a case-study of the A/B testing experimentation platform built by Netflix, a leading internet entertainment service. The innovations of this experimentation platform are interesting because they have resulted in a "technical symbiosis" of engineers and data scientists, each complementing the skill sets of the other, in order to create a platform that is robust and scalable, while also being readily extensible by data scientists.

Netflix routinely uses online A/B experiments to inform strategy and operation discussions (e.g. \cite{artwork, artwork_tech_blog, playback_ui, interleaving}), as well as whether certain product changes should be launched. Over time these discussions grew to be increasingly specialized, generating demand for more and richer metrics powered by extensible statistical methodologies that are capable of answering diverse causal effects questions.
For example, it was becoming more common for teams to require bespoke metrics to assist in the analysis of specific experiments, such as when changes to Netflix's UI architecture and video player design caused extra hard-to-isolate latency in playback startup \cite{playback_ui}; or to require bespoke statistical methodologies, such as when interleaving was used to garner additional statistical power when trying to compare two already highly-optimised personalization algorithms \cite{interleaving}.

To support these ever-growing use-cases, Netflix made a strategic bet to make their experimentation science-centric; that is, to place a heavy emphasis on supporting arbitrary scientific analyses.
To implement this science-centric vision, Netflix's experimentation platform, \texttt{Netflix XP}, was reimagined around three key tenets: trustworthiness, scalability, and inclusivity. Trustworthiness is essential since results that are untrustworthy are not actionable. Scalability is required to accommodate for Netflix's growth. Inclusivity is a key tenet because it allows scientists from diverse backgrounds such as biology, psychology, economics, mathematics, physics, computer science and other disciplines to contribute to and enrich the experimentation platform.

The implications of these tenets on \texttt{Netflix XP} are wide-ranging, but perhaps chief among them are the resulting choices of language and computing paradigm. Python was chosen as the primary language of the platform; with some components in C++ and R as needed to support performance and/or statistical models. This was a natural choice because it is familiar to many data scientists, and has a comprehensive collection of standard libraries supporting both engineering and data science use-cases. The platform also adopted a non-distributed architecture in order to reduce the barrier of entry into the platform for new statistical methodologies. Since non-distributed architectures are not as trivially scaled, the techniques employed by the platform in order to ensure scalability, i.e. compression and numerical performance optimizations, are a significant contribution of this work.

The reimagined \texttt{Netflix XP} has also had implications for its stakeholders. Firstly, data science productivity has increased. It is now straightforward for data scientists to reproduce and extend the standard analyses performed by the experimentation platform because they can run the production code in a local environment. The code also permits ad hoc extensions, allowing scientists to leverage their background and domain knowledge to easily deliver customized scorecards \cite{Fabijan2018}; for example, by including explorations of heterogenity or temporal effects. Secondly, data science workflows have been enriched with a more extensive toolkit. Since the platform was reimagined, new statistical methodologies (such as quantile bootstrapping and regression) have been contributed to the platform, which can then be used in combination with arbitrary metrics of the data scientists' choice. Thirdly, engineers have been freed up to focus on the platform itself. Since data scientists are now responsible for contributing and maintaining their own metrics and methodologies, engineers are now able to focus on aspects of the platform in which they specialize, leading to greater scalability and trustworthiness. The effect of these implications has compounded in rapid innovation cycles around ongoing strategy discussions, which has changed the face of experimentation at Netflix.

In this paper, we utilize a case-study research method to provide two main contributions. Firstly, we report on the architecture of this platform, with a special emphasis on its novel aspects: how it 
supports science-centric end-to-end workflows without compromising the engineering requirements laid out in subsequent sections.
Secondly, we describe its approach to causal inference, which leverages the potential outcomes framework to provide a unified abstraction layer for arbitrary statistical models and methodologies.

The rest of this paper is organized as follows: Section 2 presents background information in online experiments and related works. Section 3 presents the research method and validity considerations. Section 4 presents the architectural requirements, the libraries and the improvements made to \texttt{Netflix XP} to support science-centric experimentation. Section 5 discusses the causal inference framework used by \texttt{Netflix XP} that allows scientist to express their causal models in a unified way. Section 6 discusses the impact of the transition to science-centric experimentation in Netflix. Finally, Section 7 concludes the paper and discusses some future research directions.
\section{Background and Related Work}

\subsection{Online Experiments}

Online experiments have been discussed in research for over 10 years \cite{Auer2018}. The most common type of online experiment is the randomized controlled trial (RCT). RCT consists of randomly assigning users to different experience (control and treatments) of the product, while their behavior is gauged via logging a number of events. Based on this telemetry, several metrics are computed during and upon completion of an experiment. Statistical tests, such as the t-test, Mann-Whitney test, or CUPED \cite{deng2013improving} are used to identify statistically significant changes in the metrics and generate scorecards~\cite{Fabijan2018}. These scorecards help product managers, engineers, and data scientists to make informed decisions and identify a causal relationship between the product change and the observed effect. RCT in web systems is extensively discussed by Kohavi et al. \cite{Kohavi2009}. The paper presents an in-depth guide on how to run controlled experiments on web systems, discussing types of experimental designs, statistical analysis, ramp-up, the effect of robots and some architecture considerations, such as assignment methods and randomization algorithms. 

Although most research in online experiments has focused on RCT, companies have been using other types of experimental designs to infer causal relations.  For instance, Xu and Chen \cite{Chen2016} describe the usage of quasi A/B tests to evaluate the mobile app of LinkedIn. The paper details the characteristics of the mobile infrastructure that contribute to the need for designing and running different experiment designs than RCT. 

\subsection{Experimentation Processes and Platforms}

To support and democratize experimentation across multiple departments, products and use cases, Kaufman et al. \cite{kaufman_democratizing_2017} have identified the need for an experimentation platform to be generic and extensible enough to allow the design, implementation, and analysis of experiments with minimal ad hoc work. They describe, in the context of Booking.com, the usage of an extensible metric framework to provide flexibility for experiment owners to create new metrics. However, they do not describe the extensibility aspect in the context of different experimental designs and analyses as we do.

Twitter discusses its experimentation platform and how it is capable of measuring and analyzing a large number of flexible metrics \cite{DmitriyRyaboy}. The platform supports three types of metrics: built-in metrics that are tracked for all experiments, event-based metrics, and metrics that are owned and generated by engineers. One of the challenges is to scale the system with this flexibility. Scalability was achieved through several performance optimizations in their infrastructure including profiling and monitoring the capabilities in Hadoop and making processing jobs more efficient.

The trustworthiness aspect of online experiments has been an active area of research \cite{Dmitriev2017, IssaMattos2018c, Fabijan2019, Kohavi2012,zhao_online_2016}. Experiments that rely on violated assumptions or are susceptible to implementation or other design errors can lead to untrustworthy results that can compromise the conclusions and the value of the experiment. Kohavi et al. \cite{Kohavi2012} discuss lessons learned from online controlled experiments that can influence the experiment result, such as carryover effects, experiment duration, and statistical power. Fabijan et al. \cite{Fabijan2019} provide essential checklists to prevent companies from overlooking critical trustworthiness aspects of online experiments.
In our work, we do not specifically focus on trustworthiness aspects of online experiments, but on how to make the experimentation process science-centric.

More similar to our work, the different software architecture parts and design decisions of an experimentation platform are presented in Gupta et al. \cite{gupta2018anatomy}. 
The paper describes the core components of the Microsoft ExP Platform, focusing on trustworthiness and scalability. In summary, their platform can be divided into four main components: experimentation portal, experiment execution service, log processing service, and analysis service.
In the platform, experiment owners can easily create, deploy, and analyze experiments reliably and at scale. 
The platform also supports deep-dive post-experiment analysis for understanding metric changes in specific segments. However, such analysis requires a deep understanding of the structure of the data, the computation of the metrics, and the way experiment information is stored in the data.
In our work, we specifically focus on the analysis components of \texttt{Netflix XP} and describe how they have been re-designed to allow science-centric experimentation.

\section{Research Method}
This case study covers the main characteristics of \texttt{Netflix XP} that allow and support science-centric experimentation. Netflix is an entertainment media service provider and content producer with over 150 million subscribers. Within the scope of the online streaming platform, Netflix runs hundreds of experiments yearly. \texttt{Netflix XP} has been running and supporting experiments at Netflix for over 9 years. 
Over the last 3 years, an increased need for flexibility in the experimental design and analysis, as well as the need to optimize the usability of the platform for its scientists has led Netflix to redesign its experimentation infrastructure as science-centric. We conducted this case study research following the guidelines proposed by Runeson and H\"ost \cite{Runeson2009}.

\textbf{Data Collection.}
This research comprises of data collected from 2017-2019. The primary source of data consists of documentation from three regularly scheduled meetings: Experimentation Engineering with Experimentation Science leaders, Experimentation Science strategy meetings, and Experimentation Engineering with Experimentation Science verticals. Additionally, we collected data from the company-wide summit on forward thinking plans for experimentation, one-on-one interviews with data scientists, engineers, and product managers, as well as software documentation and product roadmap documents.

\textbf{Data Analysis.}
We analyzed the collected data in three steps. In the first step, we gathered all the design changes of \texttt{Netflix XP}. In the second step, we coded these changes into common groups \cite{Braun2006a}, such as requirements, architecture changes, software libraries, performance improvements, statistical methods, and causal inference modeling. Similar codes were merged, and grouped under the two main themes discussed in this paper: the software architecture and the causal inference framework.
We classified within each theme the changes that produced, and are expected to produce, high impact for the platform. We then staged the changes in a way that made the foundations of the \texttt{Netflix XP} strong.

\textbf{Validity considerations.} 
\textit{Construct validity}: Netflix has an extensive  culture of experimentation and developers and scientists regularly run online experiments. All researchers and participants were aware and familiar with the concepts, and the challenges. When needed, additional explanations and examples were given.

\textit{External validity}: This study is based on a single case company and the results and decisions taken in the architecture are dependent on this  specific context. However, the presented results can provide guidance to other organizations seeking to evolve a science-centric experimentation culture, since not all of the results and discussion are tied directly to \texttt{Netflix XP} or to the streaming service industry.

\section{Software Architecture}

Due to the ever increasing number of simultaneous experiments, experimentation platforms are often expected to derive conclusions without much human intervention. While automation brings a huge boost in velocity, 
Netflix's view is 
that it should not stand in the way of custom analyses that can leverage domain expertise in order to improve the understanding and context of the effects created by an experiment. 
Online experiments can easily become very complex and challenging to analyze \cite{Dmitriev2017}. 
In such cases, the custom designs and analyses made by the involved data scientists are of great importance.

The large number of data scientists running custom analyses required Netflix to redesign its experimentation platform. 
From Netflix's experience, when the stakeholders who need some changes are not empowered to make them, the results are sub-optimal.
When facing engineering barriers to integrate with production systems, scientists might end up creating isolated solutions that may not be integrated with the production system. 
This can lead to multiple fragmented systems with different degrees of documentation and levels of support.

This section describes the architectural components in \texttt{Netflix XP} to support science-centric experimentation. These components give data scientists full autonomy to run their analyses end-to-end, and empower them with the necessary software tools to do deep-dive analyses. We describe first the requirements of the platform and then its main components.

\subsection{Requirements for Experiment Analysis}

\begin{figure*}[t]
    \centering
    \includegraphics[width=\textwidth]{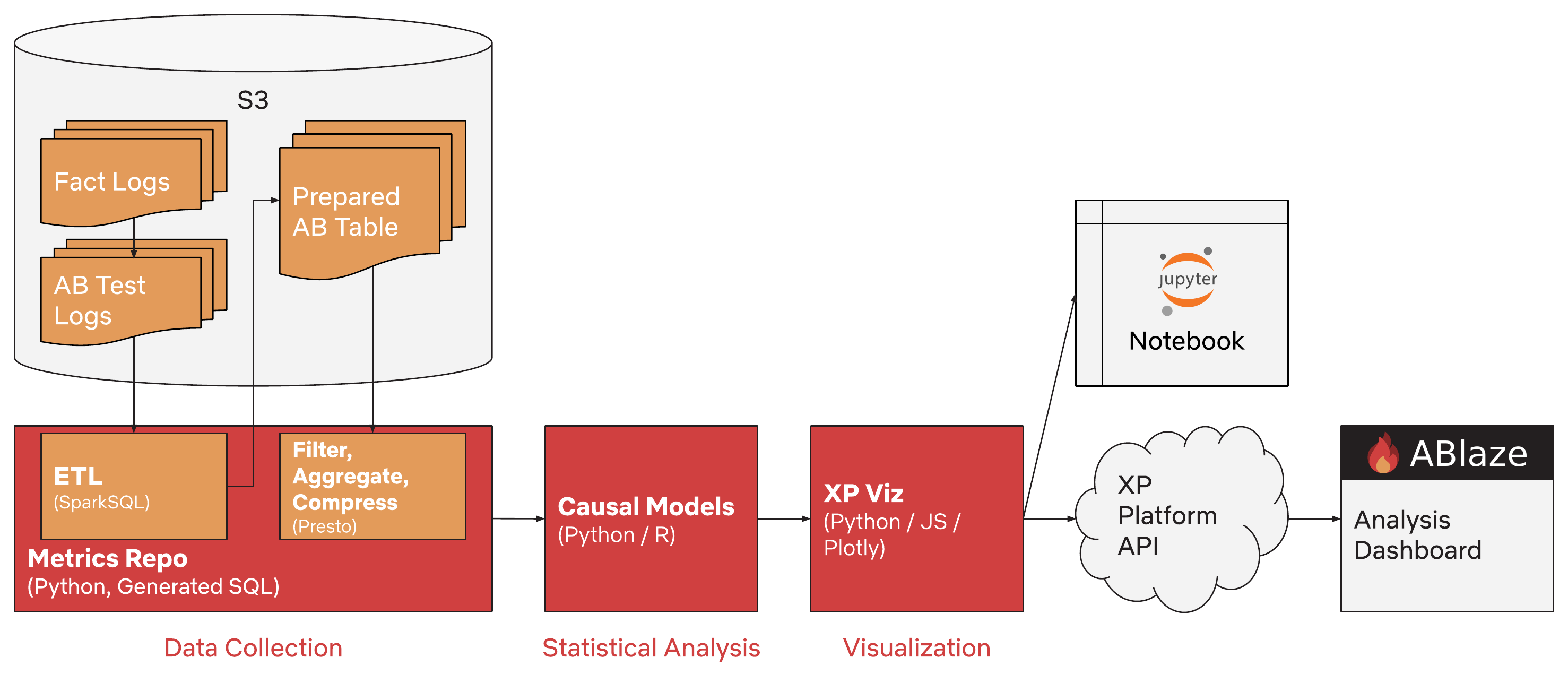}
    \caption{Experiment analysis flow at \texttt{Netflix XP}.}
    \label{fig:flow-architecture}
\end{figure*}

Regarding the analysis of online experiments, \texttt{Netflix XP} has the following requirements:

\begin{itemize}
\item \textbf{Scalable}. Each experiment at \texttt{Netflix XP} may collect and analyse data from a large portion of its 150M subscribers.
Since Netflix is a very fast growing business this
is just the starting point for the scalability requirement.
\item \textbf{Performant}. Experiment results must be calculated within seconds or minutes to allow for exploratory analyses with different user segments and metrics.
\item \textbf{Cost efficient}. The use of computational and storage resources for experiment analysis must be minimized to avoid unnecessary costs.
\item \textbf{Trustworthy}. \texttt{Netflix XP} must offer reproducible results with accurate calculations on all statistics.
\item \textbf{Usable}. Data analysts must be able to effortlessly specify standard and specialized analysis flows, view the results in an intuitive graphical interface, and perform custom deep-dive analyses.
\item \textbf{Extensible}. Data scientists from different backgrounds must be able to easily extend \texttt{Netflix XP} by contributing new experimental designs and analyses.
\end{itemize}

The last two points are directly related to science-centric experimentation.
They imply that scientists can easily setup a development environment so they can reproduce, debug, and extend analyses that happen in production. This development environment should in particular:

\begin{itemize}
    \item support interfacing with existing scientific libraries, such as Pandas, Statsmodels, and R;
    \item support local and interactive computation, for example through Jupyter Notebooks.
\end{itemize}

\subsection{Experiment Analysis Flow}
At a high level, an experiment analysis at \texttt{Netflix XP} consists of three distinct phases: data collection, statistical analysis, and visualization of the results~\cite{NetflixReimaginingExpBlogPost}. These phases along with the related components of the platform are depicted in Figure~\ref{fig:flow-architecture}.
As a first step, experiment log data are extracted, enhanced by user metadata, and stored as a table in S3. The resulting table is subsequently filtered, aggregated, and compressed based on the specified analysis configuration. 
Those first two tasks are achieved through the \texttt{Metrics Repo} component which is responsible for generating the appropriate SQL expressions that will be run on top of Spark and Presto. 
Second, different statistical methods are run on the compressed data to calculate the specified metrics of interest for the experiment. 
This step is performed by the \texttt{Causal Models} component. 
Third, different graphs are plotted for visual analysis of the results---a task of the \texttt{XP Viz} component. All of the above are orchestrated via the \texttt{XP Platform API} which is responsible for delivering the calculated metrics and produced graphs to \texttt{Netflix XP}'s frontend, \texttt{ABlaze}. Alternatively, the same analysis can also run within a Jupyter Notebook; in this case, the analyst can customize the analysis further and view the results directly in the Notebook environment. 

To enable science-centric experimentation, all three main components of \texttt{Netflix XP}'s architecture can be extended by data scientists providing new metrics, statistical methods, and visualizations. 
These three components are described in detail next.

\subsection{Metrics Repo}
\texttt{Metrics Repo} is an in-house Python framework where users define metrics as well as programmatically generated SQL queries for the data collection step. 
One of the main benefits of this is the centralization of metric definitions in a unified way.
Previously, many teams at Netflix had their own pipelines to calculate success metrics which caused a lot of fragmentation and discrepancies in calculations.

For each metric, the framework allows contributors to define certain metadata (e.g. the statistical model to use and how to be displayed). 
In order to compare a metric across two user groups, aggregate data of users in each group need to be collected and compared. 
For example, Figure~\ref{fig:metric-definition} shows the specification of a "number of streamers" metric: For each user, the number of streaming sessions with duration more than one hour are collected. For comparison between user groups, a default set of descriptive statistics, as well as proportion tests are used.

\begin{figure}
\begin{lstlisting}[linewidth=\columnwidth, breaklines=true, basicstyle=\footnotesize, language=Python]
class NumStreamers(Metric):

  def _expression(self):
    return Max(If_(self.query.streaming_hours > 0, 1, 0))

  def _statistics(self):
    return [DescriptiveStats(), ProportionStats()]
\end{lstlisting}
    \caption{Example Metric definition.}
    \label{fig:metric-definition}
\end{figure}

Many related systems in the industry generate the required data for analyses through a rigorous Extract, Transform, Load (ETL) pipeline which is responsible to annotate the available business data with the experiment data. A key design decision of \texttt{Metrics Repo} is that it moves the last mile of metric computation away from data engineering-owned ETL pipelines into dynamically generated SQL that runs on Spark. This allows scientists to add metrics and join arbitrary tables in a faster and much more flexible way since they do not have to conform to a strict predefined schema. 
The generated SQL is run only on demand and on average takes a few minutes to execute. 
This ad-hoc data collection removes the need for migrations and expensive backfills when making changes to metrics avoiding the costly and slow ETL alternative. Adding a new metric is as easy as adding a new field or joining a different table in SQL. The SQL is generated programmatically in Python which leads to a maintainable and self-documented code base. 

\subsubsection{Pre-compute vs Live-compute}
When analyzing an experiment, scientists need to see the metrics through different slicing of their data. Slicing is typically done based on different dimensions, e.g., user's country (only US users) or device type (only iOS users). 
Traditionally, to support this
statistics would be computed for each dimension value over all dimensions (pre-compute).  
Such computation leads to an explosion of possible comparisons: e.g. statistics for users in each country are compared separately, for users on different device types are again compared separately. 
The problem becomes exponential when slicing is applied via conjunction of dimension values (e.g. US users on iOS) or disjunction (e.g. users from US or Canada) due to the number of possible combinations.

To cope with the above problem, instead of pre-computing results for all the possible data slices, the platform adopts the following hybrid solution. When a new analysis is requested, statistics for only a number of commonly used slices are pre-computed. If more slices are needed, the respective statistics are computed on demand (live-compute).
Live computation is not instant, but given that on average it takes less than a minute, it is easy to queue all the different slicings and start viewing the results as they become available within seconds.
To achieve the above, the data collection is split in two steps: the first one retrieves raw data without filtering and aggregations, whereas the second retrieves the final set of filtered and aggregated data. The first part is usually much more costly to compute (multiple minutes) due to big table joins, so it is calculated in Spark and stored as a table in S3. The resulting table can subsequently be sliced with the requested filtering on demand.
For quick slicing over large amounts of data, Presto \cite{presto}, a distributed SQL engine, has been used due to its fast and interactive nature in computing filter and aggregate queries compared to alternatives such as Spark or Hive. 

Lastly, it is worth noting that, given that comparing all the pre-computed slicings is statistically controversial due to multiple hypothesis testing \cite{farcomeni2008review}, \texttt{Netflix XP} offers segment discovery through \texttt{Causal Models}, which enables automatic discovery of important data slices instead of manually comparing them one by one.

\subsubsection{Building trustworthiness.} \texttt{Metrics Repo} comes with two powerful features that increase the confidence in changes. 
The first is a testing framework which besides unit testing, it allows integration testing through which metric calculations run end-to-end on real sample data leveraging Spark. 
This enforces that every change goes through a continuous integration system ensuring that none of the well established reports are affected. 
Contributors are given the appropriate tools and are urged to follow internal best practices when submitting changes. 
The second feature is the option to run a meta-analysis on historical tests with the proposed changes. 
This enables contributors to change a metric definition and view how this would have affected hundreds of completed tests allowing them to confidently decide if they should move forward. 
Those two features have proven valuable in providing a safety net and a solid base for changes. 

\subsection{Causal Models}

\texttt{Causal Models} is a Python library that houses implementations of causal effects models and serves as the statistical engine for \texttt{Netflix XP}. Causal effects models are a restricted class of statistical models that measure causation instead of correlation, a distinction that is crucial in the context of experimentation. \texttt{Causal Models} receives data and metrics from \texttt{Metrics Repo}, then reports summary statistics such as the mean, count, and quantiles under a model, and treatment effect statistics such as the average treatment effect, its variance, its confidence interval and its p-value. Like \texttt{Metrics Repo}, the library is designed for inclusion in that it allows scientists to contribute causal effects models that integrate into the experiment analysis workflow. To support the management of many models, the \texttt{Causal Models} library also employs a modeling framework for causal inference, though we defer that discussion until section 5 and focus on \texttt{Causal Models} as a mechanism for statistical testing here.

Netflix seeks to utilize a full repertoire of causal effects models from different scientific fields in order to provide rich data for decision making.
The two-sample t-test is the most foundational causal effects model in AB testing. It is simple to understand, simple to implement, is easily scaled, and measures causal relationships instead of correlational ones when the data is randomized and controlled.
Building on top of that, ordinary least squares (OLS) is a causal effects model that can be used to determine the differences in the averages while filtering noise that the t-test cannot filter.
Quantile regression can be used to determine differences in quantiles of the distribution, for example if Netflix is concerned about changes in its most engaged users.
Panel models can be used to measure treatment effects through time.

By building modeling tools using the same stack that scientists use, \texttt{Netflix XP} was able to overcome many challenges in graduating multiple causal effects models.
Often, advances in modeling are developed by scientists with in-depth knowledge of statistics, and their methods are usually inspired by domain knowledge and experience in their field. To support their work in field experiments,
their models are developed in programming languages such as R and Python that emphasize local and interactive computing. The process of graduating such causal effects models into a production engineering system can be inefficient. First, context and knowledge must be transferred. Afterwards, the models would frequently be re-implemented in Spark in order to make them performant in a big data environment.
Implementations in a distributed computing environment, such as Spark, makes models hard to debug, and introduces a high barrier for scientists to contribute. This challenge often leads scientists to create ad-hoc applications in order to communicate their research and conclusions about an experiment. 
Instead of reimplementing models, \texttt{Causal Models} is built on Python, and engineers an interface that can integrate these models into \texttt{Netflix XP} while preserving the important engineering requirements discussed in section 4.1. 
This created a path from research directly into the experiment analysis workflow. In this case, the tenet of being inclusive to the data science stack improved the science-centric vision, as well as the tenet on trustworthiness. The innovations required to reach this milestone are discussed below.

To make contributions easier for scientists, \texttt{Causal Models} offers all the necessary support to integrate with existing statistics libraries in Python and R, the most common data science languages at Netflix. Having a multilingual framework makes \texttt{Netflix XP} inclusive to scientists from different backgrounds.
\texttt{rpy2} \cite{rpy2} has enabled the use of R inside a python framework by embedding an R process, but sharing data across them can consume large amounts of memory. In order to minimize RAM usage, the platform employs  Apache Arrow \cite{arrow}, an in-memory and cross language data format that offers zero-copy inter-process communication.
Additionally, \texttt{Causal Models} provides: (1) parallelization over multiple metrics during the calculation of statistics and (2) caching to simplify managing multiple models for multiple metrics.

Integrating with non-distributed Python and R libraries enables single-machine computation that is easy to debug and extend, however this emphasis and deviation from distributed computing can reduce the scalability of the experimentation platform. 
Therefore, the \texttt{Netflix XP} engineering team developed optimizations to scale modeling, so that the stack can still serve production and also offer local computation.
This was addressed in two ways: data compression, and high performance numerical computing. 

Data compression is an engineering achievement that allows better inclusion of the data science stack, and improves the tenet on scalability. Many causal effects models, such as OLS, compute the difference between the means of two distributions; these means are estimated using averages from a dataset. 
Some distributions can be losslessly summarized using sufficient statistics \cite{fisher1922mathematical}. 
For example, the Normal distribution can be summarized using conditional means and variances. 
When sufficient statistics are available, causal effects models do not need to be trained on the raw dataset, it can be trained on a much smaller dataset containing the sufficient statistics and features for the model. Compression rates as high as 100x were regularly observed, allowing data that would previously require hundreds of gigabytes of memory to fit in a single machine for local and interactive modeling. 
Compression is a a core part of the \texttt{Netflix XP's} analysis workflows, and is applied to all data. 
For other causal effects models that cannot be summarized using sufficient statistics, a lossy compression is used with sensible defaults that do not materially impact the precision of the treatment effect, as validated by the platform's meta-analysis framework.

Optimizing numeric computations in \texttt{Causal Models} is another way to be inclusive and performant.
At Netflix, there is a focus on high performance numerical computing applied to causal effects. 
This led to the development of causal effects primitives to support \texttt{Causal Models} through highly optimized and generic functions that are common in causal effects analysis. 
Scientists can use these primitives to compose their own analyses of experiments. 
For example, linear models are a widely used causal effects model in \texttt{Netflix XP}.
They have simple assumptions, are easy to interpret, and are highly extensible: they can be used to estimate average treatment effects, detect segments where treatment effects are different, and measure treatment effects through time. 
All of these variations can benefit from a highly optimized implementation of OLS.
Previous work from Netflix XP in \cite{linear_model_treatment_effects} demonstrates five significant optimizations to standard implementations of OLS that ultimately can compute hundreds of treatment effects over many millions of users in seconds.
Many of the causal effects primitives in \texttt{Causal Models} are developed in C++, in order to have low-level control over computation. Although scientists normally use interactive programming languages, many of their primitives are optimized in C or C++, such as the Python library NumPy \cite{van2011numpy}. 
C++ enables developers to minimize memory allocations, optimize for cache hits, vectorize functions, and manage references to data without creating deep copies, important aspects of high performance numerical computing.
Linear algebra functions that support many causal effects models are invoked through the C++ library, \texttt{Eigen} \cite{eigenweb}. 
Using C++, \texttt{Netflix XP} engineering can write optimized functions once, and deliver wrappers for python and R through \texttt{pybind11} \cite{pybind11} and \texttt{Rcpp} \cite{rcpp}, maintaining the platform's commitment to inclusivity by supporting a multilingual environment.

By creating an inclusive and scalable development experience for causal effects models, \texttt{Netflix XP} has expanded support from two-sample t-tests and Mann Whitney rank tests to many more methods, and has gained confidence that it can include more models that were not originally designed for distributed computing. 

\begin{figure}
    \centering
    \includegraphics[width=\linewidth]{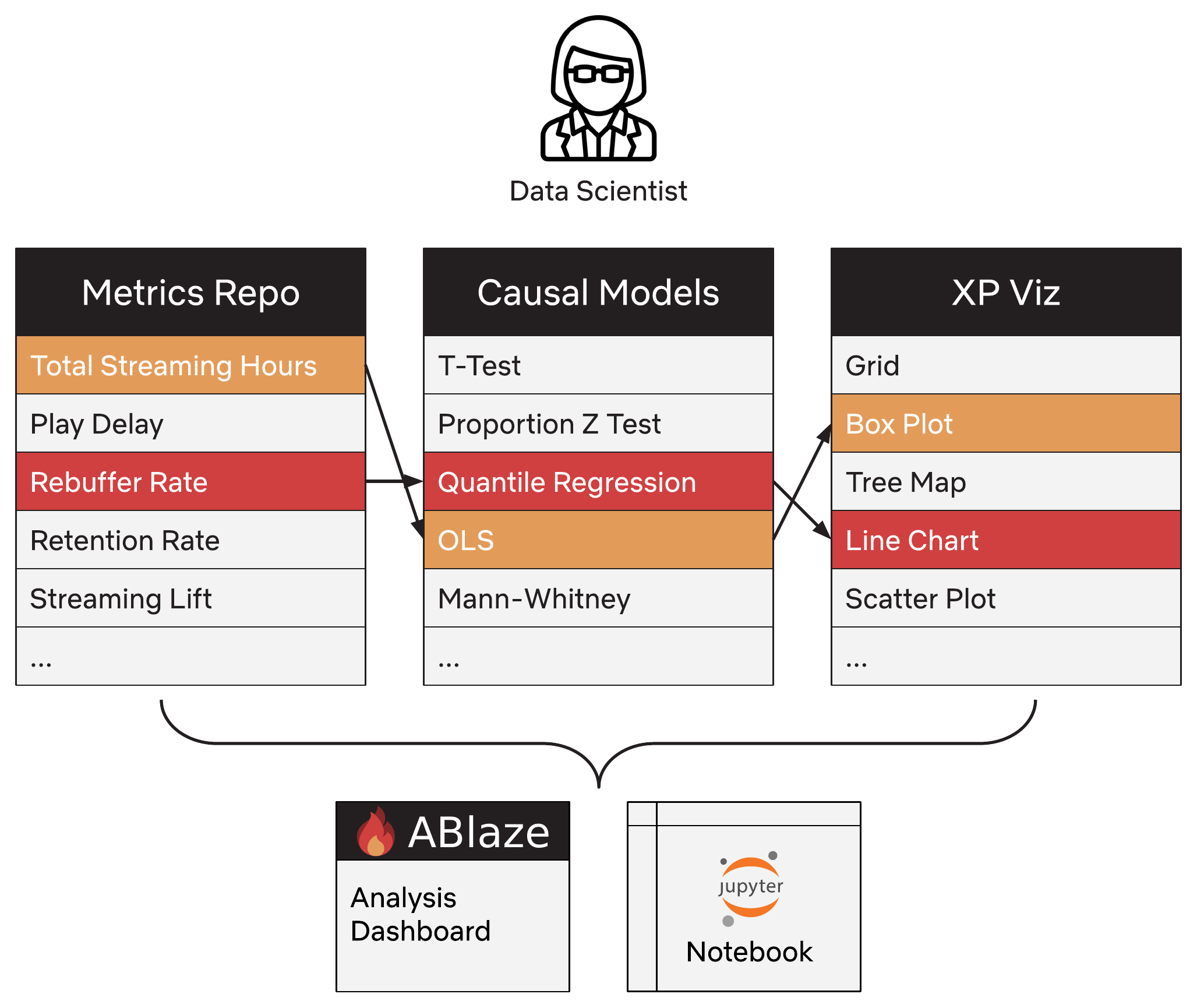}
    \caption{Examples of possible analysis flows in \texttt{Netflix XP}.}
    \label{fig:example-flow}
\end{figure}

\subsection{XP Viz}
\texttt{XP Viz} is the final component of the science-centric experimentation analysis flow in \texttt{Netflix XP}. 
It is a library that provides a lightweight and configurable translation layer from the statistical output of \texttt{Causal Models} into interactive visualizations. By implementing it as an independent pluggable component, the platform separates the view layer from the computation layer and allows reuse of standardized visualizations in other contexts. The plotting aspects of the visualization layer is based on Plotly's rich library of graphs components, allowing different teams to reuse, choose, and customize the visualizations of their metrics.

A key benefit introduced by \texttt{XP Viz} is that it provides first-class support for Jupyter Notebooks. 
Data scientists at Netflix regularly use Notebooks for their day-to-day development so supporting their familiar tooling allows them to iterate faster. 
The integration of \texttt{XP Viz} library with Notebooks allows scientists to not only compute their metrics in a Notebook when exploring, but also visualize them in the exact same way as they would do in the production UI, \texttt{ABlaze}. 
This seamless flow between the server rendered production views to local Notebook rendered views gives scientists the full power to explore and innovate using the visualizations of their choice.

\subsection{Execution of experiment analysis flows}
An analysis flow consists of metric definitions from the \texttt{Metrics Repo}, corresponding statistical tests from \texttt{Causal Models}, and visualizations from \texttt{XP Viz}. 
For instance, a possible flow, depicted in Figure~\ref{fig:example-flow}, may include the calculation of OLS on total streaming hours and visualize the results using box plots.
All of the above steps are orchestrated by \texttt{XP Platform API}, a REST API responsible for kicking off computations, keeping state, and storing results. 

One of the requirements of the \texttt{XP Platform API} is to always function in an interactive manner which means it should remain performant and with consistent latency as computational load increases. 
Such a requirement becomes more important when multiple users are interacting with the system or when a single user requests multiple slicings of an analysis. 
To achieve this, the heavy computational workload is offloaded to workers instead of using the server processes. This avoids competition for server resources as well as offers a sandboxed environment to run any potentially unsafe code. 
A common solution in such architectures is to have a list of dedicated machines that are responsible to run the jobs.
Instead, Netflix chose to to run the jobs on its implementation of OpenFaaS, a serverless computing platform.
This solution provides a lot of important features such as autoscaling in a cost effective way, efficient management of the the job queue, managed deploys as well as easy health metric and log collection. 
Leveraging OpenFaaS provides access to a cluster of machines that guarantees the interactive requirements are met as the load increases.

To unify the execution of workflows with the code in exploratory Jupyter Notebooks, a Notebook-based execution flow is enabled. 
Essentially, each execution is constructed as a parameterized Notebook that gets evaluated by the different workers. 
This Notebook can then be extracted and re-run by data scientists to fully reproduce the analysis, which allows them to debug, explore, and extend the analysis as desired. 
The described Notebook integration creates a natural cycle between the production executions and the ad-hoc Notebook explorations; a production execution can be exported to a Notebook while a Notebook execution can be promoted to production.

\section{Framework for Measuring Causal Effects}

The science-centric vision has greatly influenced the design of \texttt{Causal Models}. It offers a software framework that is not only performant, as mentioned in section 4.4, but also aligns the implementation of a causal effects model with the science of potential outcomes \cite{splawa1990application, potentialoutcomes}, making contributions from scientists natural. Potential outcomes is a generic framing of causal effects computation that is mirrored in \texttt{Causal Models}' programmatic interface. In this way \texttt{Netflix XP} is able to accomodate many causal effects models without having to worry about the domain specific implementation details of the model. 

To demonstrate how potential outcomes can be used to unify the computation of three different types of causal effects that Netflix is interested in, consider the following five statistical variables:

\begin{enumerate}
    \item $y$: The metric that needs to be measured.
    \item $X$: A binary variable indicating whether a user received treatment.
    \item $W$: Other features that are useful for modeling the variation in $y$.
    \item $t$: A variable for indexing time.
    \item $\theta$: Hyperparameters for a model.
\end{enumerate}

The potential outcomes framework is the thought exercise: what would $y$ be if we apply treatment, and what would $y$ be if we do not apply treatment? 
In a randomized and controlled experiment where all variables are constant except the treatment, any difference in $y$ must be due to noise, or to the treatment. 
Furthermore, by using this framing, a variety of treatment effects that are important for a business can be computed from an arbitrary model. 
The average treatment effect (ATE) on $y$ due to receiving the treatment, $X$, can be generically formulated as $$ATE = E[y | X = 1, W] - E[y | X = 0, W].$$ 
This treatment effect is the expected global difference between the treatment experience and the control experience. 
Likewise, the causal effect on $y$ due to $X$ for the subpopulation where $W = w^*$ is the conditional average treatment effect (CATE), and can be formulated as $$CATE(w^*) = E[y | X = 1, W = w^*] - E[y | X = 0, W = w^*].$$ 
This treatment effect shows opportunities to personalize for different segments.
In many cases, the treatment effect needs to be traced through time, which is the dynamic treatment effect $$DTE(t^*) = E[y | X = 1, W, t = t^*] - E[y | X = 0, W, t = t^*].$$
This treatment effect can show if a causal effect is diminishing, or if it can persist for a long time.
All causal effects models in \texttt{Causal Models} can subscribe to this modeling framework.

Many challenging aspects of managing causal effects models are resolved through this software abstraction based on potential outcomes.
For example, models can differ in their input, requirements, and assumptions.
A two-sample t-test accepts strictly two metrics, one for the control experience and one for the treatment, and requires that the treatment assignment was randomized. 
Ordinary least squares (OLS) accepts an arbitrary amount of metrics, the treatment assignment, and a set of covariates for the model. It requires that the treatment assignment was conditionally randomized, and that the covariates are exogeneous and full rank \cite{woolridgeIdentification}. Finally, it assumes that the noise in the metric is normally distributed. 
Both of these models assume that the observations about users are statistically independent. This assumption prevents them from being applied to time series data, where following the treatment effect through time is important; a variation of these models would have to acknowledge the autocorrelation in the data. All of these models---the t-test, OLS, and time series variations of them---have different formulas for how to determine if an effect is significant, or just noise.
Although these individual models vary, they ultimately only need to return output measuring the expected difference in the potential outcomes.

In addition to creating a path to contribute causal effects models and consolidating three types of treatment effects, \texttt{Causal Models} is able to implement the boilerplate and reduce the amount of code a scientist needs to contribute.
All three variations of treatment effects are differences in potential outcomes, where features of the model are controlled to be specific values.
They also use the same procedure: (1) train a model, (2) create a copy of the input dataset where treatment is applied to all users, (3) create another copy where treatment is not applied to any user, (4) predict the potential outcomes from each of these data copies, (5) take the average of the predictions, (6) then difference the averages. Finally, a model must implement another method for computing the variance on the treatment effect, so that it can test if the effect is significant or noise. 
This procedure is a burden to implement for every causal effects model, but it can be reduced through a simple software interface.
Each causal effects model that subscribes to the framework only needs two unique methods, then \texttt{Causal Models} completes the work that is common to how all causal effects models compute treatment effects.
The interface for an individual model requires methods to:
\begin{enumerate}
    \item \textbf{Train} a model on a dataset containing $y$ and $X$, and optionally $W$, and $t$.
    \item \textbf{Predict} the expected value of $y$ for the potential outcomes $X = 1$, and $X = 0$.
\end{enumerate}
\texttt{Causal Models} as a unifying software framework across multiple causal effects models does the work to prepare the data, invoke the train and predict methods, then aggregate and difference the output.
Optionally, a model can also implement methods for ATE, CATE, or DTE directly, for example if there is a specialized computational strategy for them, as in \cite{linear_model_treatment_effects}. Finally, the bootstrap in \cite{efron1994introduction} allows \texttt{Causal Models} to compute the variance for an arbitrary causal effects model. Developing this software framework honors the scientific study of causal effects, and is another form of engineering that can allow \texttt{Netflix XP} to better include work from scientists.

\section{Impact of Science-Centric Experimentation}

In this section, we discuss the impact science-centric experimentation has on the ability to perform deep-dive analysis and contribute new analysis flows to the platform, and reflect on the tangible results the approach brought to \texttt{Netflix XP}.

\subsection{Performing deep-dive analysis}

The architecture and framework described above make it possible for scientists to easily transition between viewing the calculated results in \texttt{ABlaze} to a deeper dive in Notebooks. 
To illustrate this flow, it is worth revising the \texttt{NumStreamers} metric example (Figure~\ref{fig:metric-definition}) to show how it can be used for further extensions and explorations. 
Assuming an analysis that includes the number of streamers metric has been calculated, from within \texttt{ABlaze} scientists can---after reviewing the results---click a button which takes them to a generated Jupyter Notebook that replicates the exact same calculations and visualizations~\cite{NetflixReimaginingExpBlogPost}. 
From there, scientists have multiple potential flows. 

First, there is the option of viewing and exploring the raw data or a reasonably sized sample. 
The data is stored as a Pandas dataframe which offers many easy ways for introspection. 
This flow is particularly useful in cases where a scientist wants to get a better sense of the actual values and their distributions in any of the segments they are interested in. 
On top of that, it is easy to join the data with other tables that were not part of the initially calculated table in order to enrich it with additional information. 
Such exploratory flows can prove of tremendous importance in analyzing tests as they provide better understanding of the data and increased insight. 

Second, scientists can alter the metric definitions and view updated calculations. 
For instance, a scientist can redefine the expression for number of streamers to be the people with at least 2 hours of viewing and re-run the statistical analysis. 
Another common use case is to explore the results of different statistical tests other than the predefined ones. 
This can be achieved by simply adding t-test or OLS in the list of statistics of the metric~(Figure~\ref{fig:metric-definition}). Lastly, a scientist can choose to visualize the results in any of the supported visualizations just by selecting any of the supported plots. 
 
\subsection{Contributing new analysis flows}
 
Within their Notebook, scientists get access to all of the source code from \texttt{Metrics Repo}, \texttt{Causal Models} and \texttt{XP Viz}. 
This allows them to edit any file they want from within the Notebook, rapidly prototype extensions, and see the impact of the changes. 
Such flow, e.g., could be used to explore the definition of a new statistical algorithm. 
All that is needed is to subclass the causal model base class and conform to the generalized causal models API. 
In a similar fashion, the definition of a new visualization can be prototyped from within a Notebook. 
Once the results are satisfactory, the extensions can be promoted back to the project by pushing the changes upstream.
This last step essentially closes the loop between \texttt{ABlaze} and Notebook since the new contributions are now available to everyone using the platform. 

\subsection{Results for Netflix}

The above features have already proven to deliver significant value in practice. 
Within less than a year of the introduction of the new architecture, more than 50 scientists have directly contributed more than 300 metrics. 
Furthermore, since the development of \texttt{Causal Models}, \texttt{Netflix XP} can broadly take advantage of OLS, quantile bootstrapping, segment discovery, quantile regression, and time series models. 
OLS, a significant foundational step, dramatically increased statistical power, making metrics more sensitive, and became a launching point for segment discovery and time series analyses.
\section{Conclusion}

In this paper we have introduced the architecture and innovations of Netflix's experimentation platform, which is routinely used to perform online A/B testing for the purposes of informing business decisions. The architecture's design was strongly influenced by a strategic bet to make the platform science-centric and support arbitrary scientific analyses on the platform, which led to its being non-distributed and written in Python. Our case-study of the platform has resulted in two novel contributions to the literature: how an experimentation platform can be designed around data science contributions without sacrificing trustworthiness and scalability, and how this is in part achieved by framing the experimentation inference problem generically enough to allow for arbitrary methodologies (in this case via the potential outcomes conceptual framework). Other innovations include compression strategies and low-level statistical optimizations that keep the non-distributed platform performant.

Since the release of the platform described in this paper, there has been a significant increase in the engagement and contributions to the experimentation platform from data scientists. This includes not only the local installation of the experimentation platform tooling, but also direct contributions of metrics and methodologies that have greatly enriched the platform and the analyses it can perform. This "technical symbiosis" of engineers and scientists has greatly increased the pace of innovation in experimentation at Netflix, and has already resulted in even deeper strategy discussions around richer analyses.

The next frontiers for the Netflix experimentation platform revolve around feature-based analyses, automation, and adaptive experiments. Feature-based analyses will allow for richer explorations of the treatment effect and interactions between multiple features in a single experiment. Automation will allow for tests to be programmatically created and modified in response to events on the platform. Adaptive experiments leverage the former two features in order to allow for automated decision making during the test; for example, this might be used to stop tests early if we have sufficient evidence \cite{gst_whitehead,gst_jennison} or use multi-arm bandits to optimally choose per-feature test allocation rates \cite{bandit_athey,bandit_zhou}. Working groups of engineers and scientists have already started collaborating on how to best approach these features in a science-centric manner.

We hope that reporting this case study will spark interest in science-centric experimentation platforms, and welcome feedback from companies or individuals interested in working or collaborating on this important topic.

\section*{Acknowledgments}
This work was partially supported by the Wallenberg Artificial Intelligence, Autonomous Systems and Software Program (WASP) funded by the Knut and Alice Wallenberg Foundation and by the Software Center.
This work has been partially sponsored by the German Ministry of Education and Research (grant no 01Is16043A) and by the Bavarian Ministry of Economic Affairs, Regional Development and Energy through the Centre Digitisation.Bavaria, as part of the Virtual Mobility World (ViM) project.

We also acknowledge Martin Tingley for supporting the science-centric vision, and Rina Chang, Pablo Lacerda de Miranda, Susie Lu, Sri Sri Perangur, and Michael Ramm for their substantial contributions to the experimentation platform.
\bibliographystyle{ACM-Reference-Format}
\bibliography{main}

\end{document}